# Laboratory characterization of SLS-based infrared detectors for precision photometry

Aaron Peterson-Greenberg[a], Michael D. Pavel[a,*]

[a]MIT Lincoln Laboratory, 244 Wood Street, Lexington, MA, 02421

**ABSTRACT**

Strained layer superlattice (SLS) detectors are a new class of infrared detectors available in the scientific and commercial markets. The photosensitive bandpass is set by material and engineered properties with typical detectors covering 7.5-10.5 microns, bluer than traditional N-band filters. SLS detectors have the potential to reach lower dark current than traditional infrared materials (like HgCdTe) allowing comparable photometric sensitivity at higher detector temperatures, easing cooling requirements. Conversely, at equal cryogenic temperatures the SLS detector will have lower dark current than HgCdTe allowing better photometric sensitivity under dark current limited operation. This work presents laboratory measurements of SLS detectors to quantify detector linearity and time stability. The potential advantages in using SLS-based detectors in future astronomical instruments is also discussed.

## 1 INTRODUCTION:

Major advances in astronomical understanding have typically been enabled by new technologies. In the infrared domain, early astronomical measurements were first made with thermopile and bolometer instruments, but a key advancement was the development of cryogenic lead sulfide devices[1]. More recently indium antimonide (InSb) and mercury cadmium telluride (HgCdTe/MCT) have been used in a variety of instruments to observe in the infrared. To avoid saturation from thermal self-emission these devices are typically cooled to cryogenic temperatures. This reduces the dark current to manageable levels, but these must still be quantified and removed during data reduction.

The future of infrared astronomical imaging will be well served by the development of improved detectors. Important metrics to consider are quantum efficiency (QE), pixel size, array format, read out noise, and dark current. A general trend with infrared InSb and MCT detector development by industry has been to increase array format and decrease pixel size while maintaining reasonable QE, dark current, and read noise[2].

One emerging infrared detector technology is strained layer superlattice (SLS). SLS utilizes engineered layers of semiconductors to fabricate infrared photodetectors. The alternating layers intentionally introduce small mismatches between the crystal structures (differences in lattice constant) to introduce controlled strain, which can suppress some dark current sources in the detector[2,3,4]. The demonstrated success of this new class of infrared detectors has led to new commercially available detectors[2,3]. These new SLS detectors could provide key benefits to infrared astronomy instrumentation in the 1-12 micron band, potentially competing with InSb and MCT detectors. Two implementations are possible: 1) using lower dark current SLS detectors in place of existing detector technologies, or 2) leveraging the lower intrinsic dark current to allow operation at higher operating temperatures and easing the cryogenic requirements on new instrumentation.

---

* Michael Pavel, E-mail: michael.pavel@ll.mit.edu

While the technology has been successfully commercialized and SLS devices have begun competing with MCT detectors in some industrial applications, the technology is still in development. Theoretical predictions for dark current suppression have not yet been realized in fabricated detectors[5]. There are also outstanding questions regarding the linearity, temporal stability of pixel response, and QE of SLS detectors that require additional study. However, SLS has the potential to become competitive with MCT detectors in precision radiometric applications and the technology should be closely followed by the scientific community.

In this paper, we report on initial laboratory testing of two commercially available SLS detectors to quantify the detector linearity and temporal stability. We begin by briefly discussing traditional MCT detectors before introducing SLS detector technology and theoretical predictions for its performance. We then outline a set of tests on two different SLS devices to characterize their ability to serve as science-grade detectors. Finally, we discuss the potential impact this technology could have on infrared astronomy if theoretical predictions are realized.

## 2 MCT DETECTORS

The dominant photodetector technology in scientific and commercial applications is bulk mercury cadmium telluride detectors. These detectors are known to have good quantum efficiency (QE), often over 90% with anti-reflection coatings[6]. Despite their popularity, MCT detectors have limitations. The cutoff wavelength of $Hg_{1-x}Cd_xTe$ detectors depends on its molar fraction/composition of mercury and cadmium, with increasing fractions of mercury lowering the bandgap and increasing the long wavelength cutoff[7,8]. However, fabrication of MCT detectors with long wavelength cutoffs are difficult to fabricate and can result in low yields from the foundry. LWIR MCT detectors can suffer from non-uniform growth defects since consistent composition of the semiconductor is difficult to maintain driving up devices cost These challenges are greater for larger format MCT detectors where material properties must be controlled over larger areas.

The limiting dark current of MCT devices can be estimated with an empirical model known as "Rule 07"[9]. This model predicts dark current in MCT detectors as a function of cut-off wavelength and detector operating temperature. Comparison with models suggests that different sources of dark current dominate as the detector cutoff wavelength changes. Detectors with cutoff wavelengths greater than 5 μm are dominated be Auger mechanisms, a process generated through the bulk of the material, and is heavily the result of band-to-band tunneling[5,6,10]. Detectors with shorter wavelength cutoffs have wider bandgaps and are likely dominated by Shockely-Read-Hall (SRH) processes[9]. Ultimately, any MCT detector will be limited by detection of thermal self-emission (sometimes called the radiative limit). In fabricated devices, this fundamental limit has generally not been reached[8,9]. Improved fabrication processes (to remove substrate defects) and cryogenic cooling have been the traditional means of reducing dark current in MCT devices.

## 3 SLS DETECTOR TECHNOLOGY

Strained layer superlattice (SLS) detectors, a recently developed technology, shows the potential to compete with MCT detectors in infrared imaging applications. These new devices could have QE similar to MCT, but with lower dark current[2,5,10,11] under the same cryogenic conditions. This could provide astronomy with a new tool for observations that would otherwise be dark current limited. From an industrial perspective, SLS is still an immature technology with significant progress in device performance metrics made over the past decade

SLS detectors were first proposed for IR detection in the 1980s. As opposed to traditional bulk detectors such as MCT, SLS detectors are heavily engineered devices. Instead of relying on composition, SLS uses the thickness of overlapping type-II band structure semiconductor layers (III-V semiconductors: typically InAs and GaSb/InAsSb layers) to determine spectral sensitivity[2,6]. Each layer acts as a well in the valence or conduction band, and simultaneously as a barrier for neighboring layers of the other material. The energy states in the wells of each band blend together to form quasi-bands, which effectively determine the absorption wavelengths[6]. The more control one has over the thickness of the SLS layers, the better one can make the device perform[2, 6, 5, 10]. This has allowed SLS devices to be tailored to a variety of cutoff wavelength from 2 to 30 μm. Special device design has even allowed dual-band detection (devices that can operate in two different band passes depending on the applied bias voltage)[6].

Recently SLS detector development has been advanced by a consortium of government and industry partners. This program created SLS detectors with large formats, high frame rates, lower noise, and multicolor functionality[6, 11]. The key breakthrough with SLS technology came from recent improvements in QE. Early SLS detectors suffered from poor QE because thin layer stacks provided SLS with a smaller active region compared to bulk MCT technology[6]. Recent process advancements have achieved QE of over 50% for substrate thicknesses greater than 3 μm[10]. Commercially produced devices generally have excellent uniformity, operability, and yield, especially when compared to MCT. Instead of traditional II-IV semiconductors, SLS use group III-V ones that have stronger and less ionic chemical bonds[2] resulting in easier manufacturability (higher yields and lower cost per unit) compared to MCT.

The structure of SLS detectors also suppresses sources of dark current that normally dominate in MCT detectors. Specifically, the biaxial compression strain on layers in SLS detectors strongly suppresses bulk Auger processes in devices with longer cutoff wavelengths[5, 10]. Figure 1 shows the dark current density of several example SLS devices at 78 K [black circles[5, 11, 12, 13]] as a function of cutoff wavelength. The predictions for MCT devices (Rule 07) is shown by the solid black line. That figure shows that SLS devices are approaching the dark currents expected for MCT devices under the same operating conditions. Although current SLS technology has yet to achieve equivalent dark current as MCT devices the technology shows promise[5, 10, 13].

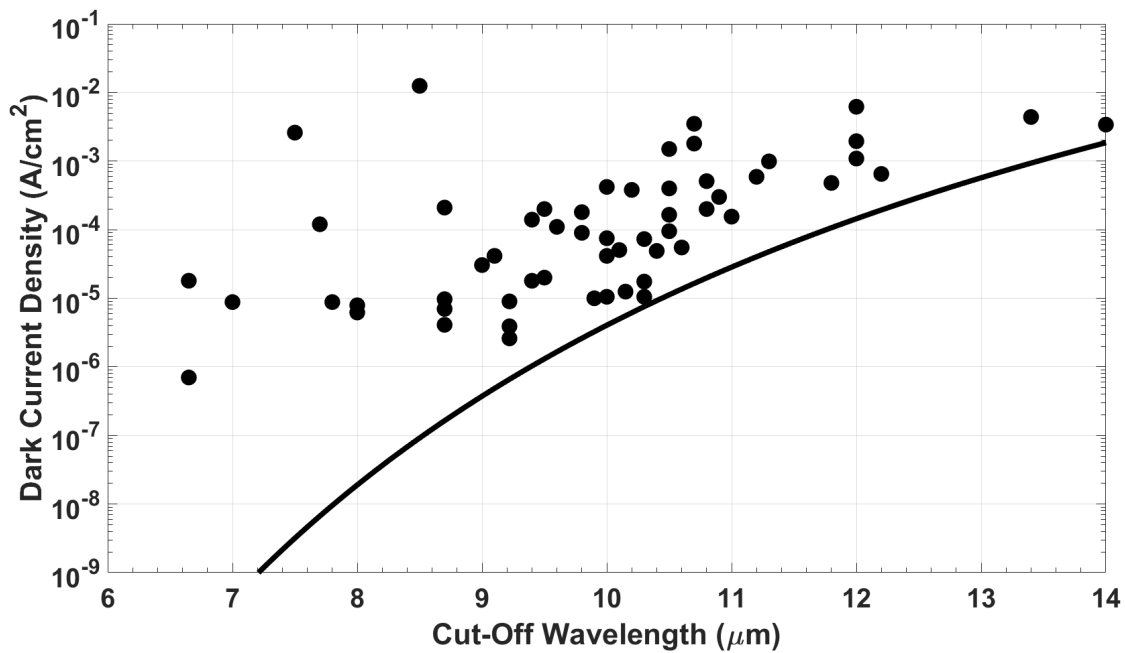

Figure 1. Example SLS detector dark-current densities (black circles) compared to MCT Rule 07 (black line). This illustrates that the dark current in SLS detectors is becoming competitive with existing MCT detectors.

The limitations to SLS detectors are now from SRH processes and high surface dark current leakage. SRH generation is caused by mid-gap defects acting as aids for thermally generated carriers. This SRH generation shortens minority carrier lifetimes, which in SLS have been found to be around 30 ns. If these defects were diminished enough to increase SRH lifetimes to the order of microseconds, SLS dark current performance would be expected to finally surpass MCT[2, 5, 10]. Figure 2 shows theoretical predictions for SLS dark current density as a function of wavelength along with the same Rule 07 prediction for MCT detectors. Black triangles show the prediction for SLS detectors with SRH lifetimes of 35 ns and black squares show the predictions for SLS if SRH processes were completely suppressed[10, 12, 14].

If SLS defects can be removed, SLS dark current would be over an order of magnitude lower than equivalent MCT devices.

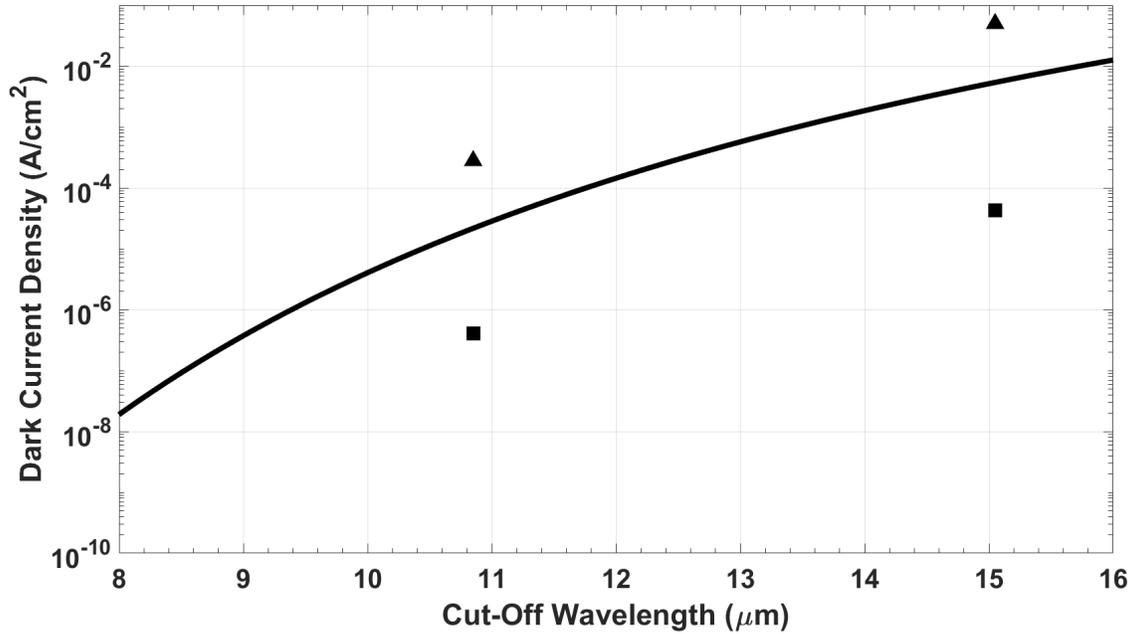

Figure 2. Theoretical SLS detector dark current densities with dominating SRH effects (black triangles) and without those effects (black squares) compared with MCT Rule 07 (black line). The potential to reduce SRH effects could lead to SLS detectors with better dark current performance than MCT.

## 4 LABORATORY RESULTS

The dark current of modern SLS detectors is approaching that of MCT devices, and potentially could outperform MCT in cryogenic application. Since SLS presents an intriguing new technology we wanted to begin quantifying several important metrics of performance. To serve as a science-grade detector any SLS device would need to demonstrate good radiometric performance. Detector linearity and temporal stability are required to calibrate an SLS instrument. A fixed number of incident photons should produce the same number of detector counts regardless of integration time or illumination intensity. Long integrations of a faint source should produce the same result as a short integration of a bright source, if the total number of incident photons is the same. Once the linearity of a detector is demonstrated, understanding the timescales upon which that linearity curve changes is important. Measuring these timescales is critical for observatory operations. An instrument that has detector calibration drifts on the order of minutes is not nearly as useful as an instrument that is stable for hours or even days. To evaluate the utility of SLS detectors for astronomical instrumentation we conducted laboratory measurements of detector linearity and temporal stability for two commercially available SLS detectors.

The setup of both tests for both devices were the same. Each detector was equipped with a 50mm lens that viewed a thermal blackbody source. The lens was intentionally defocused to prevent surface features or defects on the blackbody surface from being directly imaged. For the linearity test, the blackbody was set to a fixed temperature of 303K and the integration time of the SLS detectors was varied. For the temporal stability test, the detector was set to a fixed integration time and measurements were taken against a constant temperature blackbody over a long period of time to track variation in response.

These tests used a Model SR33-ZA blackbody from CI Systems as a uniform radiation source, operated at a constant temperature of 303 K. SLS detectors were tested from two different companies, and their properties are listed in Table 1. Device 1 was a FLIR A6750sc SLS and device 2 was a Telops FAST V1k. We note that each device has a different infrared bandpass, as listed in Table 1. Imagery was collected over a GigE connection using software provided by the vendor for each camera. All data was collected in a raw format, with all preset non-uniformity corrections in the software disabled. For both devices, a 256x256 pixel region in the center of the imager was used for all statistical analysis. A bad pixel map was measured and also applied to each device. Multiple images (1000 frames for device 1 and 256 frames for device 2) were taken under each test condition (i.e., each integration time for each test for each device) and averaged to reduce statistical noise.

Table 1. Detector parameters

| Parameter | Device 1 | Device 2 |
|---|---|---|
| Model | FLIR A6750sc SLS | Telops FAST V1k |
| Spectral Range | 7.5μm to 9.5μm | 8μm to 12μm |
| Detector Temperature | 77 K | 80 K |
| Format | 640 x 512 pixels | 640 x 512 pixels |
| Pixel Pitch | 15 μm | 25 μm |
| Min. Integration Time | 0.48 μs | 0.5 μs |
| Dynamic Range | 14 bits | 16 bits |

**4.1 Linearity**

The first quantitative test performed on each detector was a linearity calibration. A desired trait of all detectors is that the input photon flux and output detector counts are linearly proportional. For any realistic device there are known limitations to this linearity, detector bias and saturation effects, that dominate at the low and high detector count regimes, respectively. At intermediate detector count levels there should be a relatively linear response curve. We quantify the non-linear deviations as occurring when the detector response curve (flux to counts) deviates from a linear relationship by 1%.

Linearity was measured for a fixed blackbody source intensity by varying the integration times of each camera. For each camera we conducted two separate linearity measurements taken on different dates. The measured counts as a function of integration time for both devices is shown in Figure 3 (on log-log scales [left] and linear scales [right]). The results from the FLIR device are shown in the top plots and the Telops device is shown in the bottom plots. On the log-log scales the second linearity measurement (in red) has been offset by a multiplicative factors of three. The second measurement (in red) on the linear scales (right plots) has been offset vertically by 3000 detector counts. In each plot the linear regime limits are denoted by vertical gray bars and a fit to the linear data is shown by the dotted and dashed lines.

For the FLIR device, the integration time was varied from 100 nanoseconds to 1.5 milliseconds and for the Telops device the exposure time ranged from 1 to 200 microseconds. The linearity results for all measurements are similar in nature. At short integration times there is a plateau corresponding to the detector bias level and at long integration times each device is shown to saturate. A linear regime was identified in each measurement series by iterative linear regressions. For each data set a subset of manually selected data was fit and extrapolated. Measurements that differed from the linear fit by more than one percent were excluded, and points with less than one percent difference were included in the next linear regression iteration. This process converged quickly. The vertical gray bars in each panel of Figure 3 identify where the first 1% deviation was measured at the low and high integration time limits. Table 2 lists the parameters of the linear fits.

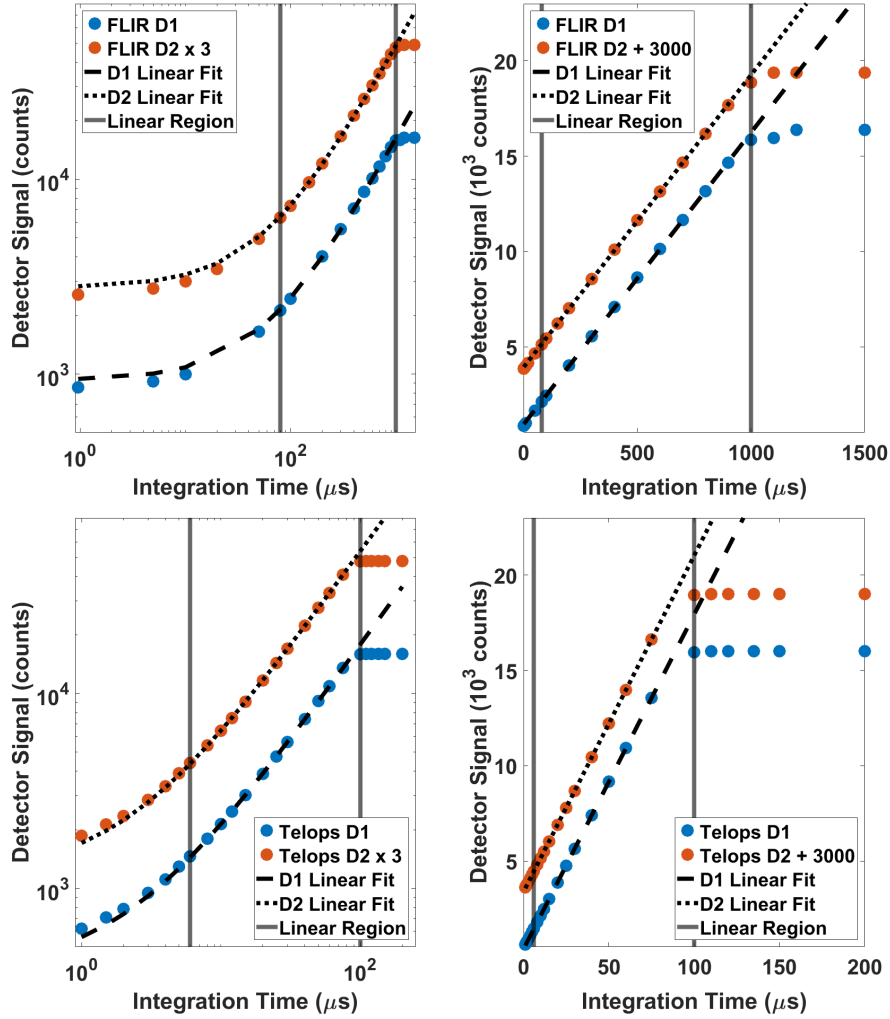

Figure 3. Detector linearity measurements of SLS detectors. [Top, left] FLIR dataset one (D1) and two (D2) with fitted linear plots (dotted and dashed lines) on a log-log scale. [Top, right] Same as top left plot except on linear axes. [Bottom, left] Telops dataset one (D1) and two (D2) with fitted linear plots (dotted and dashed lines) on a log-log scale. [Bottom, right] Same as bottom left plot except on linear axes. Vertical gray lines indicate lower and upper limits of 1% nonlinearity deviation. In each plot one measurement (red circles) has been vertically offset for clarity.

For the FLIR device (top plots in Figure 3) the linear regime extends over 11.0 dB of dynamic range, from 80 µs to 1 ms. The Telops device (bottom plots in Figure 3) has a larger linear dynamic range, 12.2 dB from 6 to 100 µs. We are unable to resolve the details of the high or low end non-linear behavior because of the limited resolution of our measurements. For comparison, the IGRINS instrument contains two science-grade MCT detectors with 1% linear dynamic ranges of 5.6 and 5.9 dB[15].

Table 2. Linearity fit parameters

| Measurement | | Slope (counts/µs) | Y-intercept (counts) |
|---|---|---|---|
| FLIR Device | Run 1 | 15.319 ± 0.089 | 928.772 ± 45.645 |
| | Run 2 | 15.339 ± 0.081 | 926.365 ± 39.334 |
| Telops Device | Run 1 | 175.504 ± 0.635 | 385.693 ± 18.534 |
| | Run 2 | 176.306 ± 0.514 | 393.476 ± 14.987 |

## 4.2 Time Stability

Temporal stability is another key property of infrared detectors. Radiometric calibration will tend to drift over time[16, 17] and this timescale determines the recalibration cadence at the telescope. Generally the response of the entire device will vary. This differs from flat fielding which measures the pixel-to-pixel response variation. To measure this timescale we measured the radiometric stability tests of both SLS detectors over the course of 4 hours. Each device was measured twice (D1 and D2) with the same blackbody source. The FLIR detector was set to an integration time of 500 μs and the Telops detector was set to 75 μs. These integration times were chosen to avoid detector saturation and were within the linear response regime.

Figure 4 shows the fraction drift in detector counts over time for each device (FLIR in top plot, Telops in bottom plot) on a logarithmic time scale up to four hours. Note the different scales for the two devices. The Telops detector had a maximum variation of about 5% from its reference point, but remained stable to with 1% drift for 30 and 50 minutes in the two measurements. The FLIR detector was found to be very stable with a maximum radiometric drift of less than 0.25% over 4 hours.

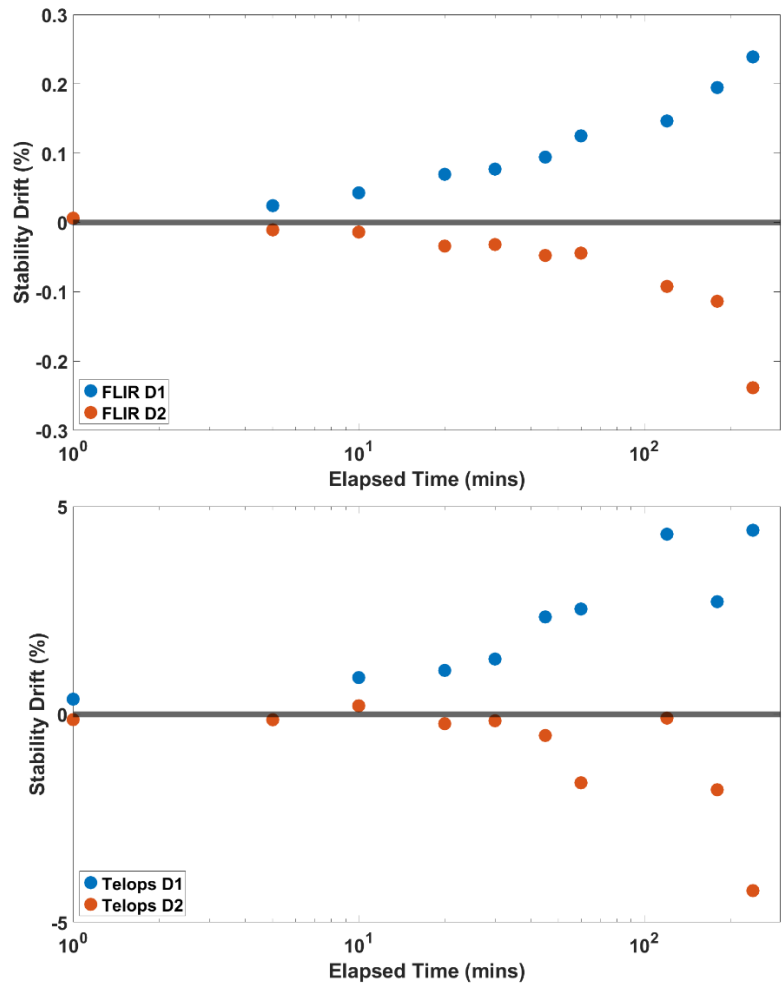

Figure 4. Temporal stability of SLS detectors shown as fractional drift in detector counts over the course of 4 hours. Two measurements for FLIR device (top plot) and Telops device (bottom plot) are shown. Note the difference in y-scales in each plot.

# 5 DISCUSSION

The key question we wanted to answer in this work was whether SLS devices had any fundamental issues that would make them unsuitable for radiometric work as a science-grade instrument. To answer this question we conducted two different measurements that quantified the responsivity of example SLS devices: linearity and temporal stability.

The first test we conducted was designed to answer if SLS detectors are fundamentally able to serve this role, and the answer seems to be yes. The linearity measurements were very promising as the devices inherently demonstrated a linear response over an order of magnitude of dynamic range. Since linearity was measured on raw detector counts, the ultimate performance of an instrument using the normal suite of detector calibration tools would likely be even better. Both SLS devices showed expected behavior at low and high count levels. The left plots in Figure 3 shows the effects of detector bias and the right plots in that same figure show saturation.

The second test was designed to quantify the temporal stability of the detector radiometric calibration. We are specifically referring to the overall drift in detector response with time, not the pixel-to-pixel variations measured by a flat field. This has implications for the actual implementation of an SLS detector in an astronomical instrument since it determines how often calibration is needed. For imaging and spectroscopy applications some standard must be observed to obtain calibrated data. This requirement is reduced when relying on differential photometry that includes a photometric standard in the appropriate bandpass. Our measurements show significant differences in the two SLS devices being tested. If we desire to limit the calibration drift to a maximum of 1% (0.01 mag) the first device must be recalibrated every 30 minutes. The second device was able to maintain its calibration to 0.25% (0.003 mag) over the four hour experiment.

Overall, either of these devices seem to be capable of science-grade operation under the right conditions. These data are not meant to suggest that any one manufacturer is better than the other. Our limited sample size prohibits generalization of these results to any form of rank ordering. There are outstanding questions regarding QE in SLS devices, beyond the scope of this work, that need additional study. Since SLS is a relatively new technology we anticipate that future generations of SLS detectors will have even better performance than demonstrated here, especially if the predictions of dark current lower than MCT detectors can be realized.

# 6 SUMMARY

SLS-based infrared detectors show promise for astronomical applications. Compared to MCT detectors, they potentially have lower dark current and could provide a new tool for observations that would already be dark-current limited. To explore this technology we have looked at two key metrics for their use in an astronomical telescope. Laboratory measurements of two different commercially-available infrared SLS detectors showed that they are inherently linear over large dynamic ranges. The temporal drift of this radiometric calibration was shown to vary considerably between the two devices tested. These tests demonstrate that current generation SLS detectors have the capability to serve as scientific sensors, but we anticipate that these characteristics will improve with time as commercially-driven fabrication produces even better devices.

# ACKNOWLEDGMENTS

This material is based upon work supported by the Department of the Air Force under Air Force Contract No. FA8702-15-D-0001. Any opinions, findings, conclusions or recommendations expressed in this material are those of the author(s) and do not necessarily reflect the views of the Department of the Air Force.